\documentclass[useAMS,usenatbib]{mn2e}

\usepackage{psfig}
\usepackage{amsmath}
\usepackage{amssymb}
\usepackage{upgreek}
\usepackage{longtable}
\usepackage[mathscr]{eucal}
\usepackage[dvips]{graphicx}
\usepackage{pdflscape}
\usepackage{array,bigstrut}

\title
[Rotation-activity in fully convective M dwarfs]
{The stellar rotation--activity relationship in fully convective M dwarfs}
\author
[Wright et al.]
{Nicholas J. Wright$^{1}$, Elisabeth R. Newton$^{2}$, Peter K.G. Williams$^{3}$, \newauthor Jeremy J. Drake$^{3}$  and Rakesh K. Yadav$^{4}$\\
\\
$^{1}$Astrophysics Group, Keele University, Keele, ST5 5BG, UK\\ e-mail: \tt{n.j.wright@keele.ac.uk}\\
$^{2}$Massachusetts Institute of Technology, 77 Massachusetts Avenue, Cambridge, MA 02139, USA\\
$^{3}$Harvard-Smithsonian Center for Astrophysics, 60 Garden Street, Cambridge, MA 02138, USA\\
$^{4}$Department of Earth and Planetary Sciences, Harvard University, 20 Oxford Street, Cambridge 02138, MA, USA\\
}

\begin{document}
\maketitle

\begin{abstract}

The coronal activity-rotation relationship is considered to be a proxy for the underlying stellar dynamo responsible for magnetic activity in solar and late-type stars. While this has been studied in considerable detail for partly-convective stars that are believed to operate an interface dynamo, it is poorly unconstrained in fully-convective stars that lack the necessary shear layer between radiative core and the convective envelope. We present new X-ray observations of 19 slowly-rotating fully-convective stars with rotation periods from the MEarth Project. We use these to calculate X-ray luminosities (or upper limits for undetected sources) and combine these with existing measurements from \citet{wrig16b}. We confirm the existence of fully-convective stars in the X-ray unsaturated regime and find that these objects follow the same rotation-activity relationship seen for partly-convective stars. We measure the power-law slope of the relationship between Rossby number (the ratio of the rotation period to the convective turnover time) and the fractional X-ray luminosity for X-ray unsaturated fully-convective stars for the first time, and find it to be consistent with that found for partly-convective stars. We discuss this implications of this result for our understanding of stellar magnetic dynamos in fully- and partly-convective stars. Finally, we also use this data to improve empirical estimates of the convective turnover time for fully-convective stars.

\end{abstract}


\section{Introduction}

Solar and late-type stars emit X-rays from a magnetically-confined plasma known as a corona \citep{vaia81}. This rarefied thermal plasma, at temperatures of several million Kelvin, was first observed on the Sun and has since been detected from all types of low-mass star. Fractional X-ray luminosities are observed at levels of $L_X / L_{bol} \sim 10^{-8} - 10^{-3}$ \citep[e.g.,][]{schm04}, with young stars typically the most active \citep[e.g.,][with $L_X / L_{bol} \sim 10^{-3}$]{tell07} and older field stars reaching down to $L_X / L_{bol} \sim 10^{-8} - 10^{-4}$ \citep[e.g.,][]{feig04}. Stellar coronae are thought to be heated by the release of magnetic energy generated by a magnetic dynamo, which itself is driven by differential rotation in the stellar interior \citep[e.g.,][]{park55}. The orders of magnitude decrease in fractional X-ray luminosity that occurs over the lifetime of a star is therefore attributed to its rotational spin-down, which is driven by mass loss through a stellar wind \citep{webe67,skum72}.

Tracers of magnetic activity, such as coronal X-ray or chromospheric H$\alpha$ emission, increase monotonically with increasing rotational velocity, a relationship first quantified by \citet{pall81} and since studied in more detail \citep[e.g.,][]{pizz03,wrig11b,rein14}. For the most rapid rotators this relationship breaks down, with X-ray luminosity saturating at log~$L_X / L_{bol} \sim -3$ \citep{vilh84,mice85}, independent of spectral type, for which the cause is currently unknown \citep{wrig11b}. When quantified in terms of the Rossby number \citep[$\mathrm{Ro} = P_{rot} / \tau$, where $\tau$ is the convective turnover time,][]{noye84} the rotation-activity relationship can be divided into saturated and unsaturated regimes at a Rossby number of $\mathrm{Ro} \sim 0.13$ \citep{wrig11b}, equivalent to rotation periods of 1--10~days for solar-type stars.

The relationship between stellar rotation and magnetic activity is a key probe of the underlying dynamo, partly because direct magnetic field observations are much harder to perform. For stars with an internal structure similar to our Sun (those with a solidly-rotating radiative core and a differentially-rotating convective envelope) this is assumed to be an $\alpha \Omega$ dynamo, first developed by \citet{park55}. The dynamo process starts with differential rotation stretching the star's poloidal field to produce a toroidal field (the $\Omega$ effect), which then rises due to magnetic buoyancy, during which turbulent helical stretching (the $\alpha$ effect) regenerates the poloidal field. In flux-transport dynamo models meridional circulation then carries the poloidal field first to the polar region of the star and then underneath the surface to regions of high shear where the process starts again \citep[though the role of meridional circulation is still debated, e.g.,][]{jouv10}. For more details of the Solar dynamo process see \citet{char10}.

Of relevance here is the location within the star where the $\Omega$-effect takes place, i.e., where strong shear stresses driven by differential rotation convert the magnetic field from a poloidal to a toroidal field. In an interface dynamo this process is believed to occur at the tachocline, the boundary between the radiative core and convective envelope \citep{dikp99}. Confirmation of the existence of such a shear layer at the interface between the solidly-rotating core and the differentially-rotating envelope has been provided through helioseismology \citep[e.g.,][]{thom96}. This region has long been thought to be a critical ingredient of an $\alpha \Omega$ dynamo, not just because it is a site of strong shear, but also because the stratification in the layer is sufficiently stable to give time for the toroidal magnetic field to be amplified before it rises to the surface. While shearing can take place within the convection zone itself, it was argued that the flux tubes created would be quickly destabilised by magnetic buoyancy \citep{park75}.

Main sequence stars with masses below $\sim$0.35~M$_\odot$ (M3.5-4) are fully convective \citep{chab97} and therefore lack the tachocline layer found in partly-convective stars. Such stars can not operate a standard Solar-type $\alpha \Omega$ dynamo. It was originally believed that fully convective stars might generate a strong but small-scale magnetic field through a turbulent dynamo \citep[e.g.,][]{durn93}. However, such objects have been observed to exhibit intense magnetic activity such as chromospheric H$\alpha$ \citep[e.g.,][]{hawl93,moha03} and coronal radio \citep{whit89} and X-ray emission \citep{wrig11b}, and strong large-scale surface magnetic fields as evidenced by Zeeman broadening \citep{john96} and spectropolarimetry \citep{mori10,see15}.

Early studies using projected rotational velocities have hinted that the rotation - chromospheric activity relationship for fully convective stars is similar to that for stars with a radiative core \citep[e.g.,][]{delf98,brow10,rein12}, though the use of projected rotation velocities prevented the form of the relationship from being quantified. Furthermore, all the fully convective stars in the large rotation period and X-ray activity sample of \citet{wrig11b} are unfortunately found in the saturated regime, which raised the question of whether such stars follow the standard rotation-activity relationship of partly-convective stars or if they only exhibit saturated X-ray emission.  Such a scenario would not be implausible if their magnetic dynamo was notably different. This issue was addressed by \citet{wrig16b} who gathered X-ray luminosities for four slowly-rotating fully convective stars and showed that they broadly follow the rotation-activity relationship known for partly-convective stars and exhibit unsaturated X-ray emission once they have sufficiently spun down. Further work by \citet{newt17} and \citet{astu17} showed that a similar rotation-activity relationship exists for chromospheric activity in fully convective stars.

In this paper we present X-ray observations of a further 19 slowly-rotating fully-convective stars to expand the sample of \citet{wrig16b}. The aim of this work is to better populate the rotation-activity diagram for fully convective stars and to constrain the form of the unsaturated regime and the relationship between rotation and X-ray luminosity. In Section~2 we present our observations and the literature data used. In Section~3 we plot these objects on a rotation-activity diagram, study the dependence between the quantities and derive new estimates of the empirical convective turnover time from the data. In Section~4 we discuss our results and their implications for dynamo theory.

\section{Observations}

Here we describe the targets selected for X-ray observations, the data reduction and analysis, and the ancillary data used.

\subsection{Target selection and stellar parameters}
\label{s-targets}

The targets for this work were primarily selected from the MEarth Project \citep{nutz08} survey of nearby fully convective M dwarfs. The proximity of the MEarth sample makes them ideal for detecting and characterising their X-ray emission. Twenty-one targets were chosen from the MEarth sample that had rotation periods measured by \citet{irwi11}, \citet{newt16} and \citet{newt18}. Periods were measured from maximum likelihood fitting and uncertainties estimated using period recovery tests. These were complemented by additional rotation periods for nearby fully convective stars from \citet{bene98}, \citet{hart11}, and \citet{kira12}.

Convective turnover times, $\tau$, were calculated using $V-K$ and the relation in \citet{wrig11b}, though since this is poorly constrained for very low-mass stars (due to the lack of unsaturated fully convective stars in that study) in Section~\ref{s-turnovertimes} we revise and improve the empirical convective turnover times for such low-mass stars. Rossby numbers, $\mathrm{Ro} = P_{rot} / \tau$ \citep{noye84}, were then calculated for all our targets. It is worth noting that the Rossby numbers of all our targets place them beyond the saturation threshold 

\begin{landscape} 
\begin{table} 
\caption{Properties of the fully convective M dwarfs studied in this work. Source names are provided from \citet[L,][]{luyt42}, \citet[G,][]{gicl71}, \citet[GJ,][]{glie79}, \citet[LSPM,][]{lepi05} and the 2MASS catalogue \citep{cutr03}. Spectral types are taken from \citet{hawl96}. Near-IR photometry are taken from 2MASS and V magnitudes primarily from UCAC4 \citep{zach13} with additional photometry from \citet{merm86}, \citet{vana95}, \citet{mone03}, \citet{jenk09} and \citet{wint11}. Rotation periods are taken from MEarth \citep{newt16,newt18} and \citet{bene98} for GJ 699 and GJ 551. Note that the rotation periods for GJ 1151, GJ 754, and 2MASS 03103891+2540535 are candidate periods. Uncertainties in $P_{rot}$ for the MEarth targets are taken from \citep{irwi11}, while for GJ~699 and GJ~551 since uncertainties are not given by \citet{bene98} we assume them to be 20\%. Convective turnover times, $\tau$, calculated using Equation 10 from \citet{wrig11b} as a function of $V-K_s$ colour.  X-ray exposure times exclude periods of high flare activity in the stars LSPM J0617+8353, GJ~1220, and LSPM J0501+2237.  For ROSAT observations the exposure times provided are those of the deep pointed observations, with X-ray fluxes calculated from the count rates provided by \citet{schm04}.  Plasma temperatures, $kT$, and X-ray fluxes are calculated from X-ray spectral fitting. X-ray fluxes are provided in both {\it Chandra}'s 0.5--8.0 keV band and ROSAT's 0.1-2.4 keV band, with $L_X / L_{bol}$ provided in the latter band, following historical standard.  Bolometric corrections were calculated from the $V-J$ colour and the relations in \citet{mann15,mann16}, allowing bolometric fluxes to be calculated for all stars, with bolometric luminosities then calculated using distances from parallaxes \citep{vana95,bene99,henr06,ried10,lepi11,ditt14,davi15,wint15,finc16,wein16,wint17}. Note that because $L_X / L_{bol}$ is independent of distance, uncertainties in the parallax do not contribute to the uncertainties on this parameter.  Upper limits are derived from the 3$\sigma$ upper bound on the source flux. }
\label{properties}
\setlength{\extrarowheight}{3pt}
\begin{tabular}{lllccccccccccc}
\hline
Name & 2MASS name & Spec. & V & J & K & $P_{rot}$ & Ro & ObsID & Exp. & $kT$ & \multicolumn{2}{c}{$f_X$ [$10^{-14}$ ergs/s/cm$^2$]} & log$_{10}$ \bigstrut[10] \\ 
\cline{13-14} & & type & [mag] & [mag] & [mag] & [days] & & & [ks] & [keV] & Chandra & ROSAT & $L_X/L_{bol}$ \bigstrut[10] \\ 
\hline 
- & J03103891+2540535 & M3.0 & 14.253  & 10.040     & 9.178 & 82.83  & 0.65 & 18948 & 3.89871 & $-$ & $< 1.02 $ & $< 1.53 $ & $< -4.50 $ \bigstrut[10] \\ 
LSPM J0617+8353 & J06170531+8353354 & M3.5 & 13.070  & 8.961     & 8.110 & 74.48  & 0.62 & 18935 & 2.24631 & $0.54^{+0.33}_{-0.21}$ & $5.32^{+2.14}_{-1.58}$ & $7.66^{+3.09}_{-2.27}$ & $-4.18^{+0.15}_{-0.16}$ \\ 
GJ 3812 & J13505181+3644168 & M3.5 & 13.709  & 9.299     & 8.422 & 72.18  & 0.53 & 18937 & 2.91931 & $-$ & $< 1.38 $ & $< 2.01 $ & $< -4.67 $ \\ 
GJ 3843 & J14211512-0107199 & M3.5 & 13.171  & 8.948     & 8.093 & 91.41  & 0.72 & 18938 & 6.85291 & $-$ & $< 0.60 $ & $< 0.87 $ & $< -5.19 $ \\ 
GJ 699 & J17574849+0441405 & M4.0 & 9.540  & 5.244     & 4.524 & 130.4  & 1.05 & \tiny{ROSAT} & 4.1 & $-$ & $9.64^{+1.17}_{-1.17}$ & $14.1^{+1.7}_{-1.7}$ & $-5.39^{+0.05}_{-0.06}$ \\ 
LSPM J0024+2626 & J00240376+2626299 & M4.0 & 14.735  & 10.222     & 9.299 & 29.84  & 0.21 & 17750 & 5.30398 & $0.58^{+0.49}_{-0.58}$ & $7.80^{+2.24}_{-1.78}$ & $11.2^{+3.2}_{-2.5}$ & $-3.49^{+0.11}_{-0.12}$ \\ 
GJ 1220 & J17311725+8205198 & M4.0 & 14.153  & 9.572     & 8.730 & 79.22  & 0.56 & 17762 & 8.87656 & $0.33^{+0.18}_{-0.33}$ & $7.61^{+1.60}_{-1.34}$ & $12.7^{+2.7}_{-2.2}$ & $-3.69^{+0.08}_{-0.09}$ \\ 
GJ 3253 & J03524169+1701056 & M4.0 & 13.855  & 8.933     & 8.053 & 81.55  & 0.56 & 14603 & 20.7010 & $0.30^{+0.05}_{-0.05}$ & $4.35^{+0.43}_{-0.39}$ & $7.76^{+0.76}_{-0.70}$ & $-4.15^{+0.04}_{-0.05}$ \\ 
GJ 3323 & J05015746-0656459 & M4.0 & 12.166  & 7.617     & 6.736 & 88.50  & 0.62 & 18940 & 1.90833 & $0.76^{+0.07}_{-0.17}$ & $55.1^{+5.3}_{-4.8}$ & $81.3^{+7.8}_{-7.1}$ & $-3.67 \pm 0.04$ \\ 
G 203-28 & J16492028+5101177 & M4.0 & 14.711  & 10.296     & 9.376 & 54.71  & 0.39 & 18942 & 5.86399 & $1.08^{+1.06}_{-0.84}$ & $0.41^{+0.40}_{-0.22}$ & $0.59^{+0.58}_{-0.32}$ & $-4.74^{+0.30}_{-0.37}$ \\ 
GJ 3417 & J06575703+6219197 & M4.5 & 13.560  & 8.585     & 7.690 & 54.50  & 0.37 & 18936 & 1.91149 & $0.38^{+0.22}_{-0.09}$ & $27.4^{+4.6}_{-4.0}$ & $43.3^{+7.3}_{-6.3}$ & $-3.54 \pm 0.07$ \\ 
GJ 1256 & J20403364+1529572 & M4.5 & 13.335  & 8.641     & 7.749 & 104.6  & 0.72 & 18939 & 5.87979 & $0.69^{+0.24}_{-0.26}$ & $2.54^{+0.75}_{-0.59}$ & $3.68^{+1.09}_{-0.86}$ & $-4.60 \pm 0.12$ \\ 
G 184-31 & J18495449+1840295 & M4.5 & 13.619  & 9.380     & 8.509 & 86.25  & 0.67 & 14604 & 7.86696 & $0.78^{+0.22}_{-0.22}$ & $2.08^{+0.41}_{-0.35}$ & $3.07^{+0.61}_{-0.52}$ & $-4.40^{+0.08}_{-0.09}$ \\ 
GJ 754 & J19204795-4533283 & M4.5 & 12.23  & 7.661     & 6.845 & 133.9  & 0.95 & 18941 & 1.91780 & $0.59^{+1.91}_{-0.28}$ & $7.73^{+3.54}_{-2.53}$ & $11.1^{+5.1}_{-3.6}$ & $-4.52^{+0.17}_{-0.18}$ \\ 
GJ 1151 & J11505787+4822395 & M4.5 & 13.25  & 8.488     & 7.637 & 117.5  & 0.81 & 18944 & 2.90040 & $-$ & $< 1.38 $ & $< 2.04 $ & $< -4.97 $ \\ 
GJ 3380 & J06022918+4951561 & M5.0 & 14.48  & 9.350     & 8.435 & 104.6  & 0.73 & 17755 & 18.8300 & $-$ & $< 0.18 $ & $< 0.27 $ & $< -5.50 $ \\ 
GJ 551 & J14294291-6240465 & M5.0 & 11.010  & 5.357     & 4.384 & 83.5  & 0.71 & \tiny{ROSAT} & 17.2 & $-$ & $222^{+2}_{-2}$ & $326^{+3}_{-3}$ & $-3.94 \pm 0.01$ \\ 
- & J19310458-0306186 & M5.0 & 16.81  & 11.147     & 10.228 & 70.92  & 0.59 & 18950 & 13.8640 & $2.16^{+2.44}_{-0.79}$ & $1.58^{+0.39}_{-0.32}$ & $1.62^{+0.40}_{-0.33}$ & $-3.93 \pm 0.10$ \\ 
GJ 1286 & J23351050-0223214 & M5.5 & 14.69  & 9.148     & 8.183 & 100.4  & 0.80 & 18946 & 2.91503 & $0.52^{+1.18}_{-0.29}$ & $1.25^{+1.23}_{-0.68}$ & $1.80^{+1.77}_{-0.98}$ & $-4.68^{+0.30}_{-0.37}$ \\ 
LSPM J2012+0112 & J20125995+0112584 & M6.0 & 15.24  & 10.486     & 9.585 & 41.15  & 0.28 & 18945 & 1.92653 & $0.67^{+0.14}_{-0.20}$ & $21.5^{+3.9}_{-3.3}$ & $31.2^{+5.6}_{-4.8}$ & $-2.93^{+0.07}_{-0.08}$ \\ 
LSPM J0501+2237 & J05011802+2237015 & M6.0 & 14.79  & 10.161     & 9.232 & 70.67  & 0.49 & 18947 & 5.01816 & $0.43^{+0.31}_{-0.18}$ & $6.83^{+1.84}_{-1.48}$ & $10.3^{+2.8}_{-2.2}$ & $-3.55 \pm 0.11$ \\ 
L 160-5 & J19194119-5955194 & - & 13.677  & 9.453     & 8.609 & 92.69  & 0.73 & 18943 & 8.93237 & $-$ & $< 0.45 $ & $< 0.66 $ & $< -5.10 $ \\ 
- & J18551471-7115026 & - & 14.668  & 10.083     & 9.222 & 90.69  & 0.64 & 18949 & 13.2540 & $-$ & $< 0.30 $ & $< 0.45 $ & $< -5.00 $ \bigstrut[10] \\ 
\hline 
\end{tabular} 
\end{table}
\end{landscape}

\noindent of $\mathrm{Ro} \sim 0.13$ \citep{wrig11b} implying that they should all be unsaturated if they follow a Solar-type activity - rotation relationship \citep{wrig16b}.

Photometry was taken from the 2-Micron All Sky Survey \citep[2MASS,][]{cutr03}, the fourth U.S. Naval Observatory CCD Astrograph Catalog \citep[UCAC4,][]{zach13}, and a number of other sources (see Table~\ref{properties}). Absolute $K$-band magnitudes were calculated using measured trigonometric (available for the majority of targets) or photometric parallaxes (see Table~\ref{properties}), and stellar masses calculated from the $M_K$ to stellar mass relation in \citet{bene16}. Bolometric corrections were calculated from the $V-J$ colour and the relations in \citet{mann15,mann16}, allowing bolometric fluxes to be calculated. Spectral types were taken from the literature (see Table~\ref{properties}).

\subsection{{\it Chandra} X-ray observations}

The targets were observed with the Advanced CCD Imaging Spectrometer \citep[ACIS,][]{garm03} on board the {\it Chandra} X-ray Observatory \citep{weis02} as part of programmes 14200934 (P.I. Wright), 17200636 (P.I. Williams) and 18200661 (P.I. Wright). The observations were performed using the ACIS-S (spectroscopy) CCD array in {\sc very faint} mode. The targets were placed on the back-illuminated S3 chip at the S-array aim point. The back-illuminated CCDs are more sensitive to soft X-rays than the front-illuminated CCDs\footnote{Note that {\it Chandra}'s back-illuminated CCDs have a slight sensitivity to visible light, but this has been calculated to be insignificant for stars fainter than 5$^{th}$ magnitude in both $I$ and $J$ bands, which is the case for all of these stars.}. The ObsIDs and exposure times are listed in Table~\ref{properties}.

The observations were processed using the CIAO~4.9 software tools \citep{frus06} and the CALDB~4.7.3 calibration files. The standard Level~1 data products were processed using the CIAO {\sc acis\_process\_events} tool to perform background cleansing and gain adjustments, with very-faint mode processing enabled. Pixel randomisation was turned off and instead the Energy-Dependent Sub-pixel Event Repositioning algorithm was enabled, which is expected to result in the most optimal event positions. A new Level~2 file was produced by filtering out events with non-zero status and bad grades (events with grades 1, 5, or 7 were removed). Intervals of high background were searched for by creating a background light curve with point sources found by {\sc wavdetect} removed. No observations showed intervals with a significant deviation from the quiescent background level and so the full observations were used for analysis.

\subsection{X-ray photon extraction}

Point source detection was performed on all observations using CIAO {\sc wavdetect}, with 14 out of the 21 targets successfully detected at the expected position. Point source extraction was performed using the {\it ACIS Extract} \citep[AE,][]{broo10} software package, which allows for positional improvement, source list refinement, spectral extraction, and the production of light curves. For the 7 sources not detected at the expected position we calculated upper limits by extracting the number of events at the target position and using the 3$\sigma$ upper bound on the source flux.

\begin{figure}
\begin{center}
\includegraphics[height=200pt, angle=270]{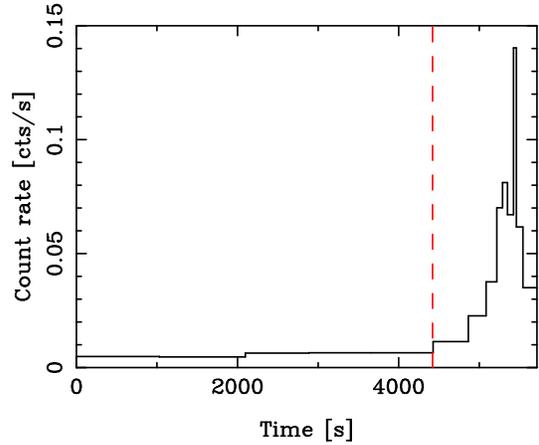}
\caption{X-ray light curve for source LSPM~J0501+2237 showing a strong X-ray flare towards the end of the observation. The red dashed line shows the time after which the data was excluded from analysis.}
\label{lightcurve}
\end{center}
\end{figure}

Low-mass stellar X-ray sources are known to show high levels of variability due to magnetic flares \citep{cara07} and rotational modulation \citep{flac05,warg17}. Bright flares can significantly increase the X-ray flux from a source, particularly during short observations such as ours, thereby inflating the quiescent flux level that would be measured. To check for this we inspected the light curves produced by AE as well as the constant-source probabilities, $P_{KS}$, derived from one-sample Kolmogorov-Smirnov tests comparing the distribution of photon arrival times with that expected for a constant source. Three sources (LSPM J0617+8353, GJ~1220, and LSPM J0501+2237) showed evidence for variability, both from their light curves and their constant-source probabilities, $P_{KS} < 0.01$. From the light curves we identified by eye the time periods during which flares contributed to the measured count rates, excluded these periods of time from the observation event files, and then re-extracted the sources. Figure~\ref{lightcurve} shows an example of one of these light curves and the time period associated with a flare that was excluded from the analysis.

\subsection{Spectral fitting and X-ray fluxes}
\label{s-spectralfits}

The extracted source spectra were fitted with thermal plasma X-ray spectral models using {\sc xspec} version 12.6.0 \citep{arna96}. The spectra were compared to {\sc apec} \citep{smit01} single-temperature models \citep{raym77} in collisional ionization equilibrium and absorbed by a hydrogen column using the {\sc tbabs} {\sc xspec} model \citep{balu92}. Due to the proximity of these objects we set the maximum hydrogen column density to be $10^{21}$~cm$^{-2}$ ($A_V \sim 0.5$, though the fitted values were much lower than this). A grid of initial thermal plasma temperatures covering $kT = 0.1-3.0$~keV was used to prevent fitting local minima, and then the model with the lowest C-statistic \citep{cash79} was selected as the best fit for each source. Only single-temperature thermal plasma models were tried for these sources as the number of photon events extracted (3 -- 130) was not high enough for more complex fits.

The results of the spectral fits, including thermal plasma temperatures and absorption-corrected broad band (0.5--8.0 keV) fluxes, are provided in Table~\ref{properties}. The uncertainties on the X-ray fluxes take into account the model fitting uncertainties as well as the uncertainties on the measured net counts. Fractional X-ray luminosities, $L_X / L_{bol}$, were calculated using the bolometric luminosities calculated earlier. The plasma temperatures are mostly in the range 0.3--0.9 keV, which are consistent with those found for other M-type dwarf stars and X-ray faint field stars \citep[e.g.,][]{mice97,wrig10b}, although one star, 2MASS~J19310458-0306, has a surprisingly high (but poorly constrained) plasma temperature of 2.2$^{+2.4}_{-0.8}$~keV. 


\subsection{ROSAT observations of additional sources}

Two objects, GJ~699 and GJ~551, were observed by ROSAT, with X-ray fluxes presented by \citet{schm04}. These have been included in our sample and their stellar and X-ray properties are presented in Table~\ref{properties}.

\subsection{Homogenisation of existing data for fully convective stars}

To facilitate reliable comparisons between the sources observed for this study and the fully convective (but rapidly-rotating) stars include in the catalogue of \citet{wrig11b} we have reprocessed those sources in the same manner as our new targets. This comprised calculating $J$-band bolometric corrections from the $V-J$ colour and the equations of \citet{mann15} for calculating bolometric luminosities and calculating stellar masses from $M_K$ and the equations of \citet{bene16}. The X-ray luminosities and rotation periods presented by \citet{wrig11b} remain unchanged.

Since the \citet{wrig11b} catalogue (and the majority of the original X-ray papers used in that work) do not provide individual X-ray flux uncertainties we adopt a standard 20\% uncertainty on X-ray flux \citep[see discussion of typical sources of uncertainty in calculating X-ray fluxes in][]{jeff11}, in addition to photometric uncertainties of 0.2~mag in $V$ \citep{hart11} and 0.1~mag for the near-IR photometry \citep{skru06}. Given that all the fully convective sources from \citet{wrig11b} have saturated X-ray emission, their uncertainties on $L_X / L_{bol}$ and Rossby number will not significantly affect our results.

\section{Rotation -- activity relationship}

\begin{figure}
\begin{center}
\includegraphics[height=220pt, angle=270]{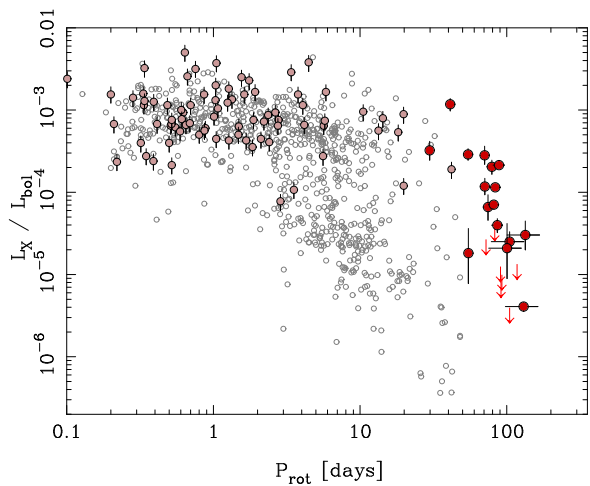}
\caption{X-ray to bolometric luminosity ratio, $L_X / L_{bol}$, plotted against the rotation period, $P_{rot}$, for the fully convective stars observed as part of this work (large red points), the fully convective stars included in the sample of \citet[][medium, light red points]{wrig11b}, and the remaining partly convective stars from that sample (grey empty circles). Error bars are shown for all fully convective stars. Upper (3$\sigma$) limits are shown for the undetected fully convective stars observed as part of this work as red arrows.}
\label{prot_relation}
\end{center}
\end{figure}

\begin{figure*}
\begin{center}
\includegraphics[height=450pt, angle=270]{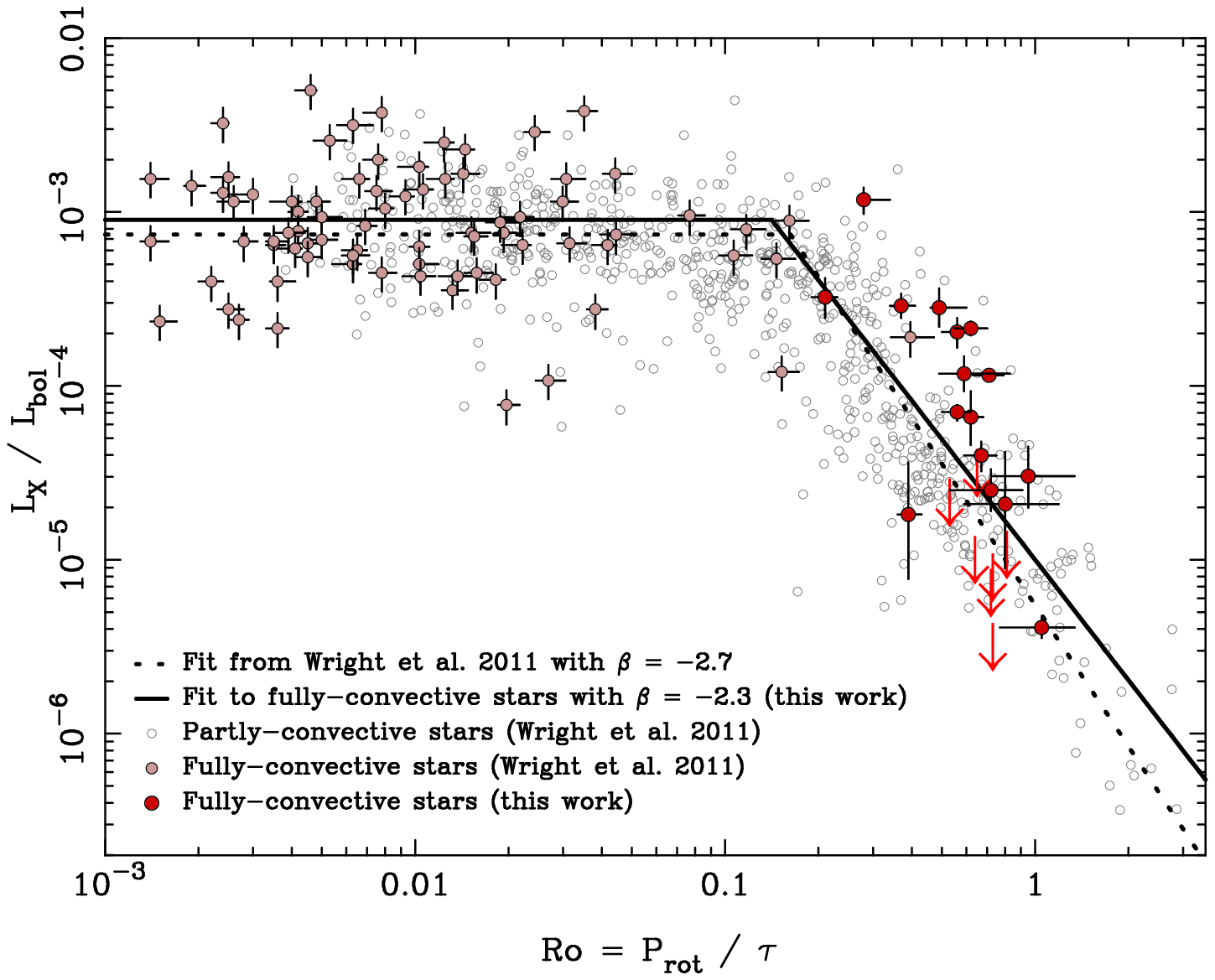}
\caption{X-ray to bolometric luminosity ratio, $L_X / L_{bol}$, plotted against the Rossby number, $\mathrm{Ro} = P_{rot} / \tau$, for the fully convective stars observed as part of this work (large red points), the fully convective stars included in the sample of \citet[][medium, light red points]{wrig11b}, and the remaining partly convective stars from that sample (grey empty circles). Error bars are shown for all fully convective stars. Upper (3$\sigma$) limits are shown for the undetected fully convective stars observed as part of this work as red arrows. The best-fitting activity - rotation relations found for fully convective stars in this work ($\beta = -2.3$ and $\mathrm{Ro}_{sat} = 0.14$, solid line) and from \citet[$\beta = -2.7$ and $\mathrm{Ro}_{sat} = 0.16$, dotted line]{wrig11b} are shown.}
\label{global_relation}
\end{center}
\end{figure*}

In this Section we study the relationship between rotation and X-ray activity in our sample of fully-convective stars. The relationship between fractional X-ray luminosity and rotation period is shown in Figure~\ref{prot_relation} and with Rossby number (the ratio of the rotation period to the convective turnover time, $\tau$) is shown in Figure~\ref{global_relation}. For the latter we have initially adopted the empirically-calibrated convective turnover times from \citet{wrig11b}, allowing us to place both partly- and fully-convective stars on the same rotation-activity diagram, and then use the apparent universality of the relationship to refine this empirical calibration. The use of empirically-determined convective turnover times does introduce a risk that they will hide interesting features in the original data (shown in Figure~\ref{prot_relation}) though we note that this will not influence the fitted slope in the unsaturated regime. This is because any change to $\tau$ will cause stars of a similar colour or spectral type to shift to lower or higher Rossby numbers by a fixed amount in log space. Since our sample of fully-convective stars does not show a correlation between colour and rotation period this shift should not affect the fitted slope, and vice versa.

\subsection{Fitting a rotation -- activity relationship for fully convective stars}
\label{s-fitting}

Figure~\ref{global_relation} shows the standard rotation -- activity diagram with the fractional X-ray luminosity, $L_X / L_{bol}$, plotted against the Rossby number, $\mathrm{Ro} = P_{rot} / \tau$ for all the fully-convective stars in our sample, both new targets and those presented by \citet{wrig11b}. The figure also shows the positions of partly-convective stars from \citet{wrig11b} that define the well-known rotation -- activity relationship: rapid rotators (those with Rossby numbers less than $\sim$0.1) show saturated activity at a level of $L_X / L_{bol} \sim 10^{-3}$, while slower rotators exhibit a power-law dependence between the fractional X-ray luminosity and the Rossby number. The new fully convective stars observed as part of this work appear to follow the same rotation-activity relationship known for partly convective stars \citep{wrig16b}.

To determine whether the slope of the rotation-activity relationship for fully-convective stars is different to that for partly-convective stars, we fit the distribution of fully-convective stars in the rotation-activity diagram with the traditional two-part power-law function commonly used for this purpose \citep[e.g.,][]{pizz03,wrig11b}. This has the form

\begin{equation}
\frac{L_X}{L_{bol}} = \left\{ \begin{array}{ll}
  C \, Ro^{\beta} & \textrm {if $\mathrm{Ro} > \mathrm{Ro}_{sat}$} \\
  \left( \frac{L_X}{L_{bol}} \right)_{sat} & \textrm{if $\mathrm{Ro} \leq \mathrm{Ro}_{sat}$}
  \end{array} \right.
\end{equation}

\noindent where $(L_X / L_{bol})_{sat}$ is the fractional X-ray luminosity for saturated stars, $\beta$ is the power-law slope for the unsaturated part of the rotation-activity relationship, $\mathrm{Ro}_{sat}$ is the Rossby number at which X-ray saturation occurs, and $C$ is a constant.

We fit this relationship using Bayesian inference and forward modelling, allowing us to model the X-ray activity levels of all the observed stars according to their rotation periods, convective turnover times, and the rotation-activity relationship in equation~1. By forward modelling our observations and fitting their measured X-ray count rates to those predicted by our model we can also include and fit the 7 undetected sources (shown in Figure~\ref{global_relation} as upper limits) based on the number of net X-ray counts calculated from source extraction performed at their expected source positions.

Our likelihood model starts by calculating Rossby numbers for our targets from their measured rotation periods and convective turnover times calculated from their $V-K_s$ colour and the relationship in \citet{wrig11b}. From this and the rotation-activity relationship in equation~1 we predict a fractional X-ray luminosity for each source. X-ray fluxes were then predicted using the bolometric luminosities as described in Section~\ref{s-targets}. For the sources observed in this study the X-ray fluxes were then converted to X-ray count rates using the {\sc xspec} model fits performed in Section~\ref{s-spectralfits} appropriate for each observation. For the fully convective stars included in the \citet{wrig11b} catalogue, for which X-ray count rates are unavailable for most sources, we performed the comparison with their X-ray fluxes.

To introduce both measurement uncertainties for measured quantities (photometry, rotation periods, and coronal plasma temperatures) as well as calibration uncertainties for inferred quantities (convective turnover times and bolometric corrections) we treat the true values of these quantities as nuisance parameters. We are effectively sampling a function, $P(\theta, \phi | x)$, that can be expanded as

\begin{equation}
P(\theta, \phi \, | \, x) = P(\theta \, | \, \phi, x) \, \, P(\phi \, | \, x)
\end{equation}

\noindent
where $\theta$ are the parameters of interest, $\phi$ are the nuisance parameters (the uncertain measurements and inferred quantities), and $x$ are the observations. We then decouple $\phi$ from its dependence on the data, assuming that $P(\phi) = P(\phi | x)$, i.e., the full observations are not informative on these nuisance parameters compared to our prior knowledge. This then reduces our function to

\begin{equation}
P(\theta, \phi \, | \, x) = P(\theta \, | \, \phi, x) \, \, P(\phi)
\end{equation}

\noindent
allowing us to marginalise out the nuisance parameters. We implement this using a two-step process following the method of \citet{lee11}, first sampling from $P(\phi)$ and then iterating several times, each time sampling from $P(\theta | \phi, x)$, before resampling from $P(\phi)$. The number of sub-iterations required to generate independent draws is typically $n = 10$ \citep{lee11}, but can be verified by inspecting the time series plots for different parameters ($\theta)$ to check there is no strong dependency on $\phi$ by the $n^{th}$ iteration.

The model has three free parameters; the logarithm of the saturation values of the fractional X-ray luminosity, log~$(L_X / L_{bol})_{sat}$, the power-law slope in the saturated regime, $\beta$, and the constant log~$C$ (that effectively sets the saturation threshold, $\mathrm{Ro}_{sat}$). A fourth parameter ($f$) was introduced to represent the scatter in the rotation-activity diagram, which may be due to a combination of inherent X-ray variability and underestimated uncertainties on various parameters \citep[see][]{hogg10}. We marginalised over $f$ to calculate uncertainties on the other fit parameters. Wide and uniform priors of $-5 < \mathrm{log} (L_X / L_{bol})_{sat} < -1$, $-10 < \beta < 0$, and $-8 < \mathrm{log} C < -1$ were used.

To sample the posterior distribution function we use the affine-invariant MCMC ensemble sampler \citep{good13} {\it emcee} \citep{fore13}, using 10,000 walkers, a total of 2000 iterations, discarding the first half as a burn-in, and a sub-iteration cadence of 10 iterations, with no evidence for a dependence between $\phi$ and $\theta$. The three parameters have very similar autocorrelation lengths with the longest being the constant $C$ of 13.9 iterations, resulting in $\sim$72 independent samples per walker. The posterior distribution functions were found to follow a normal distribution, and thus the median value was used as the best fit, with the 16$^{th}$ and 84$^{th}$ percentiles used for the 1$\sigma$ uncertainties.

The best fit slope to the unsaturated regime was found to be $\beta = -2.3^{+0.4}_{-0.6}$, which is between the canonical $\beta = -2$ value and the value of $\beta = -2.7$ found by \citet{wrig11b} from their X-ray unbiased subsample (since we have been able to include upper limits for all our undetected sources, we have no reason to think that non-detections will bias our results), though the uncertainties are sufficiently large as to be consistent with both values. This result rules out there being no dependency between Rossby number and $L_X / L_{bol}$ for fully convective stars to a confidence of $\sim$6$\sigma$. The saturation threshold for fully convective stars is estimated to be $\mathrm{Ro}_{sat} = 0.14^{+0.08}_{-0.04}$ and the saturation value is log~$(L_X / L_{bol})_{sat} = -3.05^{+0.05}_{-0.06}$, both of which are fully consistent with the values of $\mathrm{Ro}_{sat} = 0.13 \pm 0.02$ (for $\beta = -2$) or $\mathrm{Ro}_{sat} = 0.16 \pm 0.03$ (for $\beta = -2.7$), and log~$(L_X / L_{bol})_{sat} = -3.13$ found by \citet{wrig11b} for a sample predominantly composed of partly-convective stars\footnote{All the fully convective stars in the \citet{wrig11b} sample are in the saturated regime, so their fitted values of $\mathrm{Ro}_{sat}$ and $\beta$ are due entirely to partly-convective stars, making the respective fits a real comparison between partly- and fully-convective stars.}. Figure~\ref{global_relation} shows the best fit activity-rotation relationship compared to that of \citet{wrig11b}.

\subsection{Dependence of the rotation -- activity relationship on $V-K_s$ colour}
\label{s-subsets}

\begin{figure*}
\begin{center}
\includegraphics[height=450pt, angle=270]{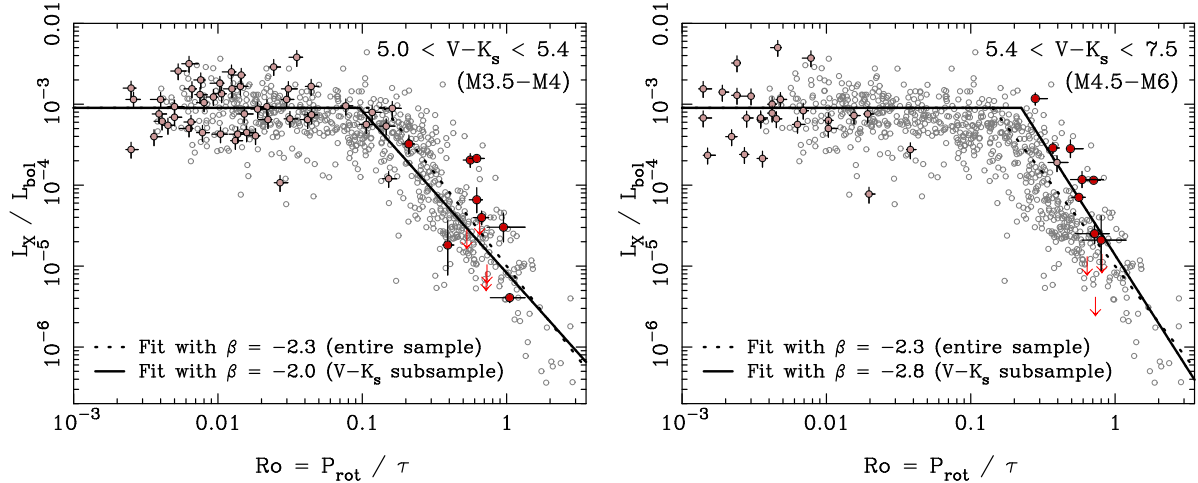}
\caption{X-ray to bolometric luminosity ratio, $L_X / L_{bol}$, plotted against the Rossby number, $\mathrm{Ro} = P_{rot} / \tau$, for two subsamples of our fully convective star sample, divided based on the $V-K_s$ colour. Symbol colour is the same as in Figure~\ref{global_relation}: fully convective stars from this work are shown as large red circles, from \citet{wrig11b} shown as medium, light red circles, and partly convective stars are shown as empty grey circles. Error bars and 3$\sigma$ upper limits are also shown. The best-fitting activity - rotation relations found for the entire sample of fully convective stars ($\beta = -2.3$, dotted line) is shown compared to the fits for each subsample, $\beta = -2.0$ (for $V-K_s < 5.4$) and $\beta = -2.8$ (for $V-K_s > 5.4$).}
\label{subsample}
\end{center}
\end{figure*}

Previous studies of the rotation-activity relationship have not found any significant variation in the slope of the unsaturated regime as a function of stellar, mass, or colour \citep[e.g.,][]{pizz03,wrig11b,newt17}. However, this is the first time that the unsaturated regime of the rotation - coronal activity relationship has been explored for fully-convective stars, and as such any variation in the slope of the unsaturated regime could provide important clues to the dynamo mechanism at work in such stars.

We divide our sample into two approximately-equal halves based on $V-K_s$ colour. These two samples, the `blue' (12 stars from our new sample with $V-K_s < 5.4$) and `red' (11 stars from our new sample with $V-K_s > 5.4$) fully-convective stars, are shown in Figure~\ref{subsample} alongside the other stars that meet these criteria from \citet[][all in the saturated regime]{wrig11b}. We then fit two-part activity-rotation relationships exactly as in Section~\ref{s-fitting}, as shown in Figure~\ref{subsample}.

We find that the two samples exhibit slopes in the unsaturated regime with power-law indexes of $\beta = -2.0^{+0.7}_{-0.4}$ (blue sample) and $\beta = -2.8^{+0.6}_{-0.8}$ (red sample), that are consistent with each other within the uncertainties (the other fit parameters are also all consistent between the fits). We thus find no significant evidence for a variation in the rotation-activity relationship for fully-convective stars of different mass or radius, although actual differences might be masked by incorrectly-calibrated convective turnover times \citep[the values we have used from][are poorly constrained for fully convective stars due to the lack of such stars in the unsaturated regime of their sample]{wrig11b}.

\subsection{Empirical determination of the convective turnover time for fully convective stars}
\label{s-turnovertimes}

While stellar convective turnover times can be calculated theoretically \citep[e.g.,][]{kim96,vent98}, it is also possible to estimate them empirically \citep[e.g.,][]{noye84,step94,pizz03,wrig11b}, based on the assumption of a universal rotation-activity relationship for active stars. However, due to the lack of previously known X-ray unsaturated fully convective stars it has been impossible to do this for fully convective stars, with earlier estimates based only on stars in the saturated regime \citep{wrig11b}. Using our data for X-ray unsaturated fully convective stars we revisit the empirical convective turnover times presented by \citet{wrig11b} and improve their estimates for fully convective stars.

To do this we follow the method of \citet{step94} and \citet{wrig11b} by fitting two-part rotation-activity relationships in the $L_X / L_{bol}$--$P_{rot}$ diagram for subsamples of stars based on their colour. The fitted relationship is of the form

\begin{equation}
\label{activity_equation}
\frac{L_X}{L_{bol}} = \left\{ \begin{array}{ll}
  C_P \, P_{rot}^{\beta} & \textrm {if $P_{rot} > P_{sat}$} \\
  \left( \frac{L_X}{L_{bol}} \right)_{sat} & \textrm{if $P_{rot} \leq P_{sat}$}
  \end{array} \right.
\end{equation}

\noindent which is exactly the same as that in equation~1, but with $P_{rot}$ instead of Rossby number as the main dependent. On the assumption of a universal activity-rotation relationship we fix the parameters $\beta = -2.7$ and log~$(L_X / L_{bol})_{sat} = -3.13$ according to the best-fitting relationship found by \citet{wrig11b} and with which our new fully convective stars are consistent. This leaves only one free parameter, the rotation period  saturation threshold, $P_{sat}$ (since the constant is given by $C_P = (L_X / L_{bol})_{sat} \, P_{sat}^{-\beta}$). As in Section~\ref{s-fitting} we fit this relationship using Bayesian inference, use the MCMC ensemble sampler {\it emcee} to sample the posterior distribution, and add an extra parameter ($f$) to represent the scatter in the relationship.

We perform these fits for colour bins of approximately the same width as those used by \citet{wrig11b} using the same sample as those authors, but complemented with the additional stars included in this study. The reddest bin was replaced with two bins to better explore how the convective turnover time varies for fully-convective stars of different colour. These bins were chosen, as in Section~\ref{s-subsets}, to divide the sample of unsaturated fully convective stars in two. The bins in $V-K_s$ colour, the number of stars in each bin, and the convective turnover times determined from our fitting process are listed in Table~\ref{results}.

\begin{table}
\begin{center}
\label{results}
\begin{tabular}{cccccc}
\hline
\multicolumn{2}{c}{$V-K_s$} & \multicolumn{2}{c}{Mass / $M_\odot$} & $N_\star$ & log $\tau$ \\
\cline{1-2} \cline{3-4}
 range & med. & range & med. & & (days) \\
\hline
1.00 -- 1.46	& 1.42	& 1.15 -- 1.36		& 1.17		& 89		& $0.93^{+0.04}_{-0.05}$			\\
1.46 -- 1.85	& 1.64	& 1.00 -- 1.14		& 1.09		& 80		& $1.12^{+0.05}_{-0.04}$			\\
1.85 -- 2.17	& 2.00	& 0.90 -- 0.99		& 0.95		& 80		& $1.17^{+0.04}_{-0.04}$			\\
2.17 -- 2.84	& 2.54	& 0.77 -- 0.89		& 0.83		& 82		& $1.36^{+0.04}_{-0.04}$			\\
2.84 -- 3.37	& 3.14	& 0.65 -- 0.76		& 0.70		& 80		& $1.42^{+0.05}_{-0.04}$			\\
3.37 -- 3.69	& 3.53	& 0.57 -- 0.64		& 0.61		& 84		& $1.52^{+0.04}_{-0.05}$			\\
3.69 -- 4.15	& 3.88	& 0.37 -- 0.56		& 0.50		& 83		& $1.72^{+0.06}_{-0.05}$			\\
4.15 -- 4.59	& 4.37	& 0.25 -- 0.36		& 0.30		& 80		& $1.82^{+0.08}_{-0.06}$			\\
4.59 -- 4.91	& 4.74	& 0.21 -- 0.24		& 0.22		& 80		& $1.96^{+0.07}_{-0.08}$			\\
4.91 -- 5.44	& 5.14	& 0.16 -- 0.20		& 0.18		& 73		& $2.07^{+0.05}_{-0.05}$			\\
5.44 -- 7.00	& 5.63	& 0.08 -- 0.15		& 0.14		& 35		& $2.19^{+0.05}_{-0.05}$			\\
\hline
\end{tabular} 
\newline
\caption{Empirical convective turnover times fitted from the combination of the data presented here and those from \citet{wrig11b}. Fits were performed under the assumption of a universal activity-rotation relationship with $\beta = -2.70$, as determined by \citet{wrig11b}. The median value of the posterior distribution was used as the best fit and the 16$^{th}$ and 84$^{th}$ percentiles used to determine the 1$\sigma$ uncertainties. Stellar masses were estimated using the empirical data in \citet{peca13} to convert $V-K_s$ to $T_{eff}$ and a 1~Gyr isochrone from \citet{sies00} to convert to stellar mass, as per \citet{wrig11b}.}
\end{center}
\end{table}

\begin{figure}
\begin{center}
\includegraphics[height=220pt, angle=270]{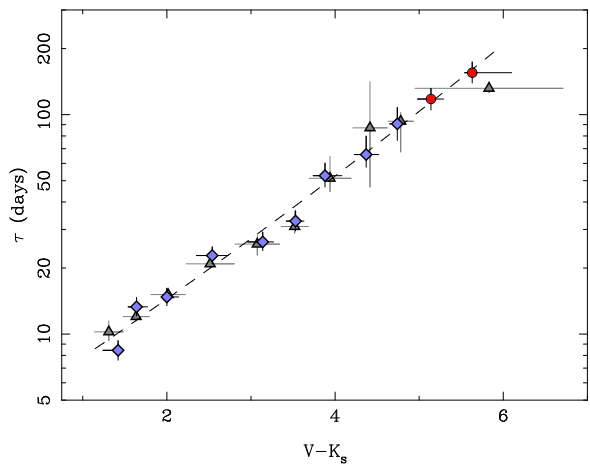}
\caption{Empirical convective turnover time as a function of $V-K_s$ colour using the best-fitting activity-rotation model parameters determined by \citet{wrig11b} with $\beta = -2.70$ and log~$(L_X / L_{bol})_{sat} = -3.13$. Error bars (1$\sigma$) are shown for all points, which for $V-K_s$ colour are the 16$^{th}$-84$^{th}$ percentiles of the colour distribution within each bin. Grey triangles show the convective turnover times determined by \citet{wrig11b}, while blue diamonds and red circles show the values determined in this work for partly-convective stars \citep[data entirely from][]{wrig11b} and for fully-convective stars (data from that paper and this work), respectively. The slight differences between the results of this work and \citet{wrig11b} for partly-convective stars can be attributed to the different fitting techniques employed. The dashed line shows the best-fitting polynomial relationship between the two quantities.}
\label{tau}
\end{center}
\end{figure}

Figure~\ref{tau} shows the best-fitting convective turnover times determined as a function of $V-K_s$ colour and compared to those determined previously by \citet{wrig11b}. Our estimates of the convective turnover times for partly-convective stars differ slightly from those determined by \citet{wrig11b}, which can be attributed to slight differences in the fitting technique (we fixed the $L_X / L_{bol}$ saturation level and introduced a free parameter to account for the large scatter in the rotation-activity relationship, which those authors did not). For fully convective stars we see that the convective turnover time continues to rise, with no evidence for the flattening-off hinted at in the results of \citet{wrig11b}. Note that the slightly lower convective turnover time we find at the blue end may not be real as it may be caused by the onset of super-saturation for rapidly-rotating F-type stars \citep[see Figure~6 of ][]{wrig11b}.

We parameterise these results by fitting various polynomial functions to log~$\tau$ as a function of $V-K_s$. We fit these functions using MCMC, varying $V-K_s$ for each fit in accordance with the distribution of values within each bin. We find a good fit with a simple straight line, with no evidence that either higher-order polynomials or multi-part functions are necessary for a better fit. The best fit is

\begin{equation}
\mathrm{log} \tau = 0.64^{+0.10}_{-0.12} + 0.25^{+0.08}_{-0.07} (V-K_s)
\end{equation}

\noindent which is valid over the range $1.1 < V-K_s < 7.0$ and has an rms dispersion in log~$\tau$ of 0.045~dex. This parameterisation of $\tau$ should replace the equivalent fit in \citet{wrig11b}, which was highly uncertain for $V-K_s > 5$ due to their lack of unsaturated fully-convective stars. Converting the $V-K_s$ colours to $T_{eff}$ using the data from \citet{peca13} and converting these to stellar mass using a 1~Gyr iscohrone from \citet{sies00} we can also parameterise $\tau$ in terms of stellar mass (see Table~\ref{results}) and fit a polynomial relationship of the form

\begin{equation}
\mathrm{log} \tau = 2.33^{+0.06}_{-0.05} - 1.50^{+0.21}_{-0.20} (M / M_\odot) + 0.31^{+0.16}_{-0.17} (M / M_\odot)^2
\end{equation}

\noindent which is valid over the range $0.08 < M / M_\odot < 1.36$ and has an rms dispersion in log~$\tau$ of 0.055~dex. More complex functions were explored but did not provide a better fit.

\section{Discussion}

We have measured X-ray luminosities for an expanded sample of slowly-rotating, fully-convective stars so that they can be plotted in an X-ray activity - rotation diagram to study correlations between the two quantities. We find that the sources follow the classical rotation -- activity relationship known for partly-convective stars and hinted at in the study of \citet{wrig16b}, with a power-law relationship between Rossby number and fractional X-ray luminosity for slow rotators and a saturated fractional X-ray luminosity for fast rotators. With our expanded sample we are able to quantify the power-law slope in the unsaturated regime for the first time and fit a value of $\beta = -2.3^{+0.4}_{-0.6}$, consistent with that found for partly convective stars of $\beta = -2.70 \pm 0.13$ by \citet{wrig11b}. There is a hint that the slope may vary as a function of $V-K_s$ colour for the fully-convective stars, but this is equally-well explained by (and probably more likely due to) variations in the convective turnover time.

Since the rotation - activity relationship is believed to be a proxy for the underlying magnetic dynamo, this similarity between the form of the relationships for partly- and fully-convective stars is surprising because the two types of star have long been thought to operate very different magnetic dynamos \citep[e.g.,][]{durn93}. As discussed by \citet{wrig16b} the most likely explanations for this are either that fully convective stars operate the same dynamo to that found in partly-convective stars like the Sun (which would therefore require that the tachocline is not a vital ingredient for such a dynamo) or that fully convective stars generate a purely turbulent dynamo (one that does not rely on a shear layer) that exhibits a rotation-activity relationship very similar to that in partly-convective stars.

Turbulent dynamos are those that do not rely on a shear layer and are sometimes referred to as $\alpha^2$ dynamos. \citet{durn93} were among the first to explore their potential, finding that they would produce a small-scale magnetic field in fully-convective stars, but that a large-scale field could not be generated without a shear layer. Such models were only weakly dependent on rotation and could not provide the rotation-activity relationship we see in Figure~\ref{global_relation}. Later 3D dynamo simulations by \citet{dobl06} and \citet{brow08} were able to generate larger-scale magnetic fields with some evidence that rotation may influence the resulting magnetic field strength. More recent 3D dynamo simulations of M-type stars without a tachocline have been able to reproduce both cyclic behaviour on 5--15~year timescales \citep{fan14}, the detailed small-scale and large-scale features of magnetic fields in fully convective stars \citep{yada15}, and decreases in field strength and emergence of magnetic cycles at slow rotation periods \citep{yada16}. The latter feature is in line with recent reporting of activity cycles in late-type stars without a tachocline \citep{rout16,warg17}. While such a dynamo could produce the observed trend between rotation and X-ray activity that we have observed, it is perhaps surprising that two very different dynamo mechanisms would have the same relationship between these two quantities.

An alternative hypothesis is that fully-convective stars actually operate a dynamo very similar to that found in our Sun. \citet{spru11} has argued that both observations and simple theoretical considerations point to the solar convection zone as the only necessary ingredient for dynamo action. The Solar dynamo requires a layer of strong shear to induce the $\Omega$ effect that converts the poloidal field to a toroidal field. Regions of strong shear do exist within stellar convection zones, but since magnetic flux tubes are highly buoyant it has been suggested that this would not provide sufficient time for the magnetic field to be amplified to sufficient strength before it rises to the surface \citep{park75}. At the bottom of the convection zone the tachocline provides a layer of strong shear where magnetic buoyancy is weak and as such the field can be sufficiently amplified before it reaches the stellar surface. However, current 3D dynamo simulations have found that differential rotation within the convection zone of stars can generate large-scale toroidal magnetic fields through the $\Omega$ effect without the need for a tachocline \citep[e.g.,][]{muno09,brow10,nels13,yada15,yada16}. These {\it distributed} dynamo models may suggest that previous {\it interface} dynamo models, where the toroidal field is generated just at the tachocline, are not necessary.

Most dynamo models are not at the stage where they can predict correlations between rotation and either magnetic field strength or magnetic activity indicators such as X-ray activity. As theory progresses observational constraints such as the form of the rotation-activity relationship for fully-convective stars are vital for identifying the key features such dynamos should exhibit and how they differ (or not) to those in solar-like stars.

\section{Conclusions}

We have combined new X-ray observations of 19 slowly-rotating fully-convective stars with existing X-ray data and rotation periods from various sources. From the resulting X-ray luminosities and upper limits we have produced the first X-ray activity-rotation diagram for fully convective stars that is well populated in both the saturated and unsaturated regimes. We confirm the results of \citet{wrig16b} that fully-convective stars exhibit a rotation-activity relationship that is indistinguishable from that of partly-convective stars. We measure the power-law slope of the relationship between Rossby number and fractional X-ray luminosity for X-ray unsaturated fully-convective stars for the first time, and find it to be consistent with that measured for partly-convective stars. We then use this data to improve the empirical estimation of the convective turnover time for fully-convective stars, which was previously poorly constrained.

Our results suggest that either fully-convective stars operate a turbulent dynamo that leads to a rotation-activity relationship with the same dependence on Rossby number as that for partly-convective stars, or that both types of star operate very similar dynamos that rely on the interaction of rotation and turbulent convection. We suggest that the latter explanation is more likely, especially in the light of recent models that have shown that $\alpha \Omega$ dynamos can operate without the aid of a tachocline, with the shearing action distributed throughout the convective envelope instead of being concentrated in the interface between the radiative core and the convective envelope.

\section{Acknowledgments}

We thank the anonymous referee for a careful and helpful reading of the paper as well as comments and suggestions that have improved the quality of this paper. NJW acknowledges an STFC Ernest Rutherford Fellowship (grant number ST/M005569/1). ERN is supported by an NSF Astronomy and Astrophysics Postdoctoral Fellowship under award AST-1602597. JJD was supported by NASA contract NAS8-03060 to the {\it Chandra X-ray Center} and thanks the director, H. Tananbaum, and the science team for continuing support and advice. This research has made use of data from the {\it Chandra X-ray Observatory} and software provided by the {\it Chandra X-ray Center} in the application packages CIAO and Sherpa, and from Penn State for the {\sc ACIS Extract} software package. This research has also made use of NASA's Astrophysics Data System and the Simbad and VizieR databases, operated at CDS, Strasbourg.

\bibliographystyle{mn2e}
\bibliography{/Users/nwright/Documents/Work/tex_papers/bibliography.bib}
\bsp

\end{document}